\documentclass[12pt]{article}

\usepackage{afterpage,amsfonts,amsmath,amssymb,array}
\usepackage{epsfig,hhline,longtable}
\usepackage{tabularx}
\usepackage{algorithm,algorithmic}

\newlength{\defbaselineskip}
\newcommand{\setlinespacing}[1]%
           {\setlength{\baselineskip}{#1 \defbaselineskip}}

  \def\t0{{t_0}}

\allowdisplaybreaks
\begin{document}
\thispagestyle{empty}
%\vspace*{1.5cm}
%
\begin{center}
{\LARGE \bf Exact Simulation of the 3/2 Model}
\end{center}

\vspace*{1.0cm}

\begin{center}

{\large
\renewcommand{\thefootnote}{\arabic{footnote}}
{\bf Jan Baldeaux}\footnote{University of Technology Sydney,
School of Finance $\&$ Economics, PO Box 123, Broadway, NSW,
2007, Australia } } \vspace*{0.5cm}

\today

\end{center}

%\vspace*{10.0cm}
\vspace*{1.5cm}

{\small
\begin{center}
\vspace*{0.5cm}
\begin{minipage}[t]{12cm}
{\bf Abstract.}
This paper discusses the exact simulation of the stock price process underlying the $3/2$ model. Using a result derived by Craddock and Lennox using Lie Symmetry Analysis, we adapt the Broadie-Kaya algorithm for the simulation of affine processes to the $3/2$ model. We also discuss variance reduction techniques and find that conditional Monte Carlo techniques combined with quasi-Monte Carlo point sets result in significant variance reductions.
\\[25mm]
\end{minipage}
\end{center}}
%
%\noindent %{\small
% {\em JEL Classification:\/}  \\
%1991 {\em Mathematics Subject Classification:\/} primary ;\\
%secondary . \\
 \vspace*{2cm}

\setlinespacing{1.66}

\newpage

\section{Introduction}

Exact simulation allows us to sample solutions of stochastic differential equations (SDEs) from the appropriate distribution functions. Alternatively, one could discretize the time interval and simulate the solution by stepping through the time grid. The latter approach has two drawbacks. Firstly, a bias is introduced, on which it is often difficult to obtain a priori estimates. Secondly, the time discretization usually increases the computational complexity of the simulation. Regarding the topic of time discretizations for SDEs, we refer the reader to \cite{KloedenPl95}, and for methods which show how to trade-off bias and variance, we refer the reader to \cite{DuffieGl95}, and also the recent work on multilevel methods, see for example \cite{CreutzigDeMuRi09} and \cite{Giles08}. Exact simulation methods enjoy the appealing properties of avoiding a simulation bias and the increased computational complexity due to time-stepping. Finally, they also allow us to asses the quality of discretization schemes.

For some SDEs, the correct distribution of the solution is well-known or easy to obtain, such as Brownian motion and its direct transformations, such as geometric Brownian motion and the Ornstein-Uhlenbeck process. Furthermore, for processes in the family of the squared Bessel process, such as the Bessel and the square-root process, exact simulation schemes are also known, see \cite{RevuzYo99} and \cite{PlatenHe06}. Recently, the exact simulation of general diffusion processes has become topical, see for example the works \cite{BeskosRo05}, \cite{BeskosPaRo06}, \cite{BeskosPaRo08}, and \cite{Chen08}.

In a related paper, \cite{BroadieKa06} solved an important problem of both practical and theoretical interest, namely the exact simulation of stochastic volatility and affine jump-diffusion processes, in particular, the Heston model and its extensions. Roughly speaking, the method developed in \cite{BroadieKa06} can be described as follows: One firstly samples the integrated variance conditional on which the stock price follows a lognormal distribution, whose variance parameter is determined by the integrated variance. To be more precise, \cite{BroadieKa06} first sample the variance at the final time point, using the known distribution of the solution of the square-root process, and consequently derive the Laplace transform of the conditional distribution of the integrated variance, relying on a result from \cite{PitmanYo82}.

In this paper, we adopt an analogous approach: we also sample the variance at the final time point first, and consequently derive the Laplace transform of the conditional distribution of the integrated variance. The technique is analogous to the one employed by \cite{BroadieKa06} and is found in \cite{CraddockLe09}, where Lie Symmetry Methods are employed to derive Laplace transforms of functionals of diffusions, such as squared Bessel processes. Having obtained the Laplace transform of the distribution, we have reduced the problem to sampling from the lognormal distribution, which is trivial.

The $3/2$ model was introduced in \cite{Heston97}, and also studied in \cite{Lewis00}, \cite{CarrSu07}, and \cite{ItkinCa10}, and is empirically supported, see \cite{ItkinCa10}. It is a stochastic volatility model, and is related to the Heston model, \cite{Heston93}, in the following way: Under the Heston model, the variance process is modeled via a square-root process, under the $3/2$ model, the variance process is modeled via the inverse of a square-root process. We point out that the variance process underlying the $3/2$ model, the inverse of a square-root process, has also been used to model interest rates, see e.g. \cite{AhnGa99}.

From a mathematical point of view, the problem is also interesting: though not an affine process, the $3/2$ model is still analytically tractable, in particular, the characteristic function of the logarithm of the stock price still has a closed-form solution. Invoking a result obtained via Lie Symmetry Analysis, which also allows us to recover the result presented in \cite{PitmanYo82}, we manage to obtain the Laplace transform of the conditional distribution of the integrated variance. This emphasizes that results from Lie Symmetry Analysis can also be employed when designing Monte Carlo methods, whereas so far the main application of Lie Symmetry Analysis to finance has been in the derivation of closed-form pricing formulae, see e.g. \cite{LaurenceWa05}, \cite{CraddockLe07}.

We also discuss variance reduction techniques to speed up the simulation. Since \cite{Willard97}, conditional Monte Carlo techniques have been known to improve the efficiency of Monte Carlo algorithms, especially when combined with quasi-Monte Carlo methods, see also \cite{BroadieKa06}. The results we present confirm this observation: we manage to achieve considerable variance reductions.

The remainder of the paper is structured as follows. In Section \ref{secMonteCarloalg}, we state the algorithm we employ to simulate the stock price process under the $3/2$ model. The main challenge of the algorithm lies in the simulation of the conditional integrated variance, which is discussed in Section \ref{secdistribconintegratedvar}. In Section \ref{secimplementation}, we detail the implementation of the algorithm and show some numerical results. These results are improved on in Section \ref{secvarred}, where variance reduction techniques are studied. Section \ref{secconc} concludes the paper.

\section{The Monte Carlo Algorithm} \label{secMonteCarloalg}

The dynamics of the stock price under the $3/2$ model under the risk-neutral measure are given by
\begin{eqnarray} \label{eqstock}
\frac{d S_t}{S_t} &=& r dt + \sqrt{V_t} \rho dW^1_t + \sqrt{V_t} \sqrt{1- \rho^2} dW^2_t \, ,
\\ \label{eqvariance} dV_t &=& \kappa V_t \left( \theta - V_t  \right) dt + \epsilon V^{3/2}_t d W^1_t \, .
\end{eqnarray}
Equation \eqref{eqstock} describes the dynamics of the stock price, and Eq. \eqref{eqvariance} the dynamics of the variance process. We denote by $W^1$ and $W^2$ two independent Brownian motions. Regarding the parameters, $r$ represents the constant interest rate, $\rho$ the instantaneous correlation between the return and the variance process and $\epsilon$ governs the volatility of volatility. The speed of mean reversion is given by $\kappa V_t$ and $\theta$ denotes the long-run mean of the variance process. However $V=\left\{ V_t \, , \, t \geq 0 \right\}$ is just the inverse of a square-root process, as we now demonstrate, in particular, we introduce the square-root process $X=\left\{ X_t \, , \, t \geq 0 \right\}$ via $X_t = \frac{1}{V_t}$. The dynamics of $X_t$ are given by
\begin{displaymath}
d X_t = \left( \kappa + \epsilon^2 - \kappa \theta X_t \right) dt - \epsilon \sqrt{X_t} d W^1_t \, .
\end{displaymath}
%\begin{eqnarray*}
%dX_t &=& \frac{-dV_t}{V^2_t} + \frac{1}{V^3_t} d \left[ V \right]_t
%\\ &=& - \frac{\kappa}{V_t} \left( \theta - V_t \right) dt - \epsilon V^{-1/2}_t dW^1_t + \epsilon^2 dt
%\\ &=& \left( \kappa - \frac{\kappa \theta}{V_t} \right) dt - \epsilon V^{-1/2}_t d W^1_t + \epsilon^2 dt
%\\ &=& \left( \kappa + \epsilon^2 - \frac{\kappa \theta}{V_t} \right) dt - \epsilon V^{-1/2}_t d W^1_t
%\\ &=& \left( \kappa + \epsilon^2 - \kappa \theta X_t \right) dt - \epsilon \sqrt{X_t} d W^1_t \, .
%\end{eqnarray*}
Hence, using the process $X$, we obtain the following dynamics for the stock price, where $u>t$,
\begin{eqnarray*}
S_u &=& S_t \exp \left\{ r (u-t) - \frac{1}{2} \int^u_t \left( X_s \right)^{-1} ds + \rho \int^u_t \left( \sqrt{X_s} \right)^{-1} dW^1_s  \right\}
\\ && \exp \left\{  \sqrt{1- \rho^2} \int^u_t \left( \sqrt{X_s} \right)^{-1} dW^2_s  \right\} \, .
\end{eqnarray*}
We aim to adopt the approach from \cite{BroadieKa06}. In this regard, it is useful to study $\log(X_t)$, for which we obtain the dynamics
\begin{displaymath}
d \log(X_t)= \left( \frac{ \kappa + \frac{\epsilon^2}{2} }{X_t} - \kappa \theta  \right) dt - \epsilon \left( \sqrt{X_t} \right)^{-1} dW^1_t \, .
\end{displaymath}
%\begin{eqnarray*}
%d \log(X_t) &=& \frac{dX_t}{X_t} - \frac{d [X]_t}{X^2_t}
%\\ &=& \frac{1}{X_t} \left( \kappa +\epsilon^2 - \kappa \theta X_t \right) dt - \epsilon \left( \sqrt{X_t} \right)^{-1} dW^1_t - \frac{1}{2} \frac{\epsilon^2}{X_t} dt
%\\ &=& \left( \frac{\kappa +\epsilon^2}{X_t} - \kappa \theta - \frac{1}{2} \frac{\epsilon^2}{X_t} \right) dt - \epsilon \left( \sqrt{X_t} \right)^{-1} dW^1_t
%\\ &=& \left( \frac{ \kappa + \frac{\epsilon^2}{2} }{X_t} - \kappa \theta  \right) dt - \epsilon \left( \sqrt{X_t} \right)^{-1} dW^1_t \, .
%\end{eqnarray*}
Hence
\begin{displaymath}
\log (X_u) = \log(X_t) + \left( \kappa + \frac{\epsilon^2}{2} \right) \int^u_t \frac{ds}{X_s}  - \kappa \theta (u - t) - \epsilon \int^u_t \left( \sqrt{X_s} \right)^{-1} dW^1_s \, ,
\end{displaymath}
or equivalently
\begin{equation} \label{eqBMintegral}
\int^u_t \left( \sqrt{X_s} \right)^{-1} dW^1_s = \frac{1}{\epsilon} \left( \log \left( \frac{X_t}{X_u} \right) + \left( \kappa + \frac{\epsilon^2}{2} \right) \int^u_t \frac{ds}{X_s}   - \kappa \theta (u - t) \right) \, .
\end{equation}

We hence arrive at Algorithm \ref{alg:3over2}, which is analogous to the Broadie-Kaya algorithm from \cite{BroadieKa06}.

\begin{algorithm}[H]
\caption{Exact Simulation Algorithm for the $3/2$ model}\label{alg:3over2}
\begin{algorithmic}
\STATE \STATE \textbf{Step 1)} Simulate $X_u \bigg| X_t$ using the
noncentral $\chi^2$-distribution \STATE \textbf{Step 2)} Simulate
$\int^u_t \frac{ds}{X_s} \bigg| X_t , X_u$ \STATE \textbf{Step 3)}
Recover $\int^u_t \left( \sqrt{X_s} \right)^{-1} dW^1_s$ from Eq.
\eqref{eqBMintegral} \STATE \textbf{Step 4)} Simulate $S_u$ given
$S_t$, $\int^u_t \left( \sqrt{X_s} \right)^{-1} dW^1_s$ and
$\int^u_t \left( X_s \right)^{-1} ds$ via
    \begin{eqnarray*}
    \lefteqn{ \log (S_u) }
    \\ &\sim& N \left( \log (S_t) + r(u-t)  - \frac{1}{2} \int^u_t \left( X_s \right)^{-1} ds + \rho \int^u_t \left( \sqrt{X_s} \right)^{-1} dW^1_s , \sigma^2(t,u) \right) \, ,
    \end{eqnarray*} where
    \begin{displaymath}
    \sigma^2(t,u) = \left( 1 - \rho^2 \right) \int^u_t X^{-1}_s ds  \, .
    \end{displaymath}
\end{algorithmic}
\end{algorithm}

%Following \citeN{BroadieKa06}, we have the following algorithm
%\begin{itemize}
%\item simulate $X_u \bigg| X_t$ using the noncentral $\chi^2$-distribution,
%\item simulate $\int^u_t \frac{1}{X_s} ds \bigg| X_t , X_u$,
%\item recover $\int^u_t \left( \sqrt{X_s} \right)^{-1} dW^1_s$ from equation \eqref{eqBMintegral},
%\item simulate $S_u$ given $S_t$, $\int^u_t \left( \sqrt{X_s} \right)^{-1} dW^1_s$ and $\int^u_t \left( X_s \right)^{-1} ds$ from
%    \begin{eqnarray*}
%    \lefteqn{ \log (S_u) }
%    \\ &\sim& N \left( \log (S_t) + r(u-t)  - \frac{1}{2} \int^u_t \left( X_s \right)^{-1} ds + \rho \int^u_t \left( \sqrt{X_s} \right)^{-1} dW^1_s , \sigma(t,u) \right) \, ,
%    \end{eqnarray*} where
%    \begin{displaymath}
%    \sigma^2(t,u) = \left( 1 - \rho^2 \right) \int^u_t X^{-1}_s ds  \, .
%    \end{displaymath}
%\end{itemize}
Regarding Step 1) of Algorithm \ref{alg:3over2}, from the dynamics of $X$ it is clear, see e.g. \cite{JeanblancYoCh09}, that conditional on $X_t$,
\begin{displaymath}
\frac{X_u \exp \left\{ \kappa \theta (u-t) \right\}}{c(u-t)} \sim \chi^2 \left( \delta, \alpha \right) \, ,
\end{displaymath}
where $\delta= \frac{4 (\kappa + \epsilon^2)}{\epsilon^2}$, $\alpha=\frac{X_t}{c(u-t)}$, and
\begin{displaymath}
c(t)=\epsilon^2 \left( \exp \left\{ \kappa \theta t \right\} - 1 \right)/(4 \kappa \theta) \, .
\end{displaymath}
Step 2) is discussed in Section \ref{secdistribconintegratedvar}, in fact, we derive the Laplace transform of the conditional distribution of
\begin{displaymath}
\int^u_t \frac{ds}{X_s} \bigg| X_t, X_u  \, .
\end{displaymath}
Steps 3) and 4) are trivial.

\section{The Distribution of the Conditional Integrated Variance} \label{secdistribconintegratedvar}

We now discuss the simulation of $\int^u_t \frac{ds}{X_s}  \bigg|
X_t, X_u$, where we follow the approach from \cite{BroadieKa06}.
Recall that
\begin{displaymath}
X_t = X_0 + \int^t_0 \kappa \theta \left( \frac{\kappa + \epsilon^2}{\kappa \theta } - X_s \right) ds + \int^t_0 \left( - \epsilon \right) \sqrt{X_s} dW^1_s \, .
\end{displaymath}

We firstly change the constant in front of $\sqrt{X_s}$ to $2$:
\begin{eqnarray*}
\int^t_0 \left( - \epsilon \right) \sqrt{X_s} dW^1_s &=& 2 \int^t_0 \sqrt{X_s} \left( \frac{\epsilon}{2} \right) \left( - dW^1_s \right)
\\ &\stackrel{\textrm{law}}{=}& 2 \int^t_0 \sqrt{X_s} dW^1_{\frac{\epsilon^2 s}{4}}
\\ &=& 2 \int^t_0 \sqrt{X_{\left( \frac{4 \epsilon^2 s}{\epsilon^2 4} \right)}} d W^1_{\frac{\epsilon^2 s}{4}} \, .
\end{eqnarray*}
Now, setting $u = \frac{\epsilon^2 s }{4}$, we get $du = \frac{\epsilon^2}{4} ds$, and hence
\begin{eqnarray*}
X_{\left( \frac{4 \epsilon^2 t}{\epsilon^2 4} \right)} = X_0 + \frac{4}{\epsilon^2} \int^{\frac{\epsilon^2 t}{4}}_0 \kappa \theta \left( \frac{\kappa +\epsilon^2}{\kappa \theta} - X_{\frac{4 u }{\epsilon^2}} \right) du + 2 \int^{\frac{\epsilon^2 t}{4}}_0 \sqrt{X_{\frac{4 u}{\epsilon^2}}} dW^1_u \, .
\end{eqnarray*}
Defining $\xi(u) = X_{\frac{4 u}{\epsilon^2}}$ we arrive at
\begin{displaymath}
\xi (\frac{\epsilon^2 t}{4}) = \xi(0) + \frac{4}{\epsilon^2} \int^{\frac{\epsilon^2 t}{4}}_0 \kappa \theta \left( \frac{\kappa +\epsilon^2}{\kappa \theta} - \xi(u) \right) du + 2 \int^{ \frac{\epsilon^2 t}{4}}_0 \sqrt{\xi(u)} dW^1_u \, .
\end{displaymath}

We now introduce $n=\frac{4 \kappa \theta ( \kappa + \epsilon^2)}{ \epsilon^2 \kappa \theta }$ and $j= - \frac{2 \kappa \theta}{\epsilon^2}$, then
\begin{equation} \label{eqrhotincludingj}
\xi(t) = \xi(0) + \int^t_0 \left( 2 j \xi(u) + n \right) du + 2 \int^t_0 \sqrt{\xi(u)}dW^1_u \, .
\end{equation}
To develop a formula for
\begin{displaymath}
E \left( \exp \left\{ - a \int^t_0 \frac{ds}{X_s}   \right\}
\bigg| X_t, X_u  \right)  \, , \end{displaymath} we firstly apply
a change of law formula to eliminate the random component of the
drift term and get a process with $j=0$ in
\eqref{eqrhotincludingj}, i.e. a squared Bessel process. Again, we
proceed as in \cite{BroadieKa06}.
\begin{eqnarray*}
\lefteqn{E \left( \exp \left\{ - a \int^u_t \frac{ds}{X_s}
\right\} \bigg| X_t , X_u \right)}
\\ &=& E \left( \exp \left\{ - a \int^u_t \frac{ds}{\xi \left( \frac{\epsilon^2 s}{4}  \right)}   \right\} \bigg| \xi \left( \frac{\epsilon^2 t}{4} \right), \xi \left( \frac{\epsilon^2 u}{4} \right) \right)
\\ &=& E \left( \exp \left\{ - \frac{4 a}{\epsilon^2} \int^{\frac{u \epsilon^2}{4}}_{\frac{t \epsilon^2}{4}}  \frac{dl}{\xi_l}  \right\} \bigg| \xi \left( \frac{\epsilon^2 t}{4} \right) , \xi \left( \frac{\epsilon^2 u}{4} \right) \right)
\\ &=& \tilde{E} \left( \exp \left\{ - \frac{4a}{\epsilon^2} \int^{\frac{u \epsilon^2}{4}}_{\frac{t \epsilon^2}{4}} \frac{dl}{\xi(l)} - \frac{j^2}{2} \int^{\frac{u \epsilon^2}{4}}_{\frac{\epsilon^2 t}{4}} \xi(l) dl \right\} \bigg| \xi \left( \frac{\epsilon^2 t}{4} \right) , \xi \left( \frac{\epsilon^2 u}{4} \right) \right) /
\\ && \tilde{E} \left( \exp \left\{ - \frac{j^2}{2} \int^{\frac{\epsilon^2 u}{4}}_{\frac{\epsilon^2 t}{4}} \xi(l) dl \right\} \bigg| \xi \left( \frac{\epsilon^2 t}{4}  \right) , \xi \left( \frac{\epsilon^2 u}{4} \right) \right) \, ,
\end{eqnarray*}
where the last equality follows from formula (6.d) in \cite{PitmanYo82} and $\tilde{E}$ denotes the expectation taken with respect to the law of the squared Bessel process. So we need formulae for
\begin{equation} \label{eqcondLPtransform1}
\tilde{E} \left( \exp \left\{ - \frac{4a}{\epsilon^2}
\int^{\frac{u \epsilon^2}{4}}_{\frac{t \epsilon^2}{4}}
\frac{dl}{\xi(l)} - \frac{j^2}{2} \int^{\frac{u
\epsilon^2}{4}}_{\frac{\epsilon^2 t}{4}} \xi(l) dl \right\} \bigg|
\xi \left( \frac{\epsilon^2 t}{4} \right) , \xi \left(
\frac{\epsilon^2 u}{4} \right) \right)
\end{equation}
and
\begin{equation} \label{eqcondLPtransform2}
\tilde{E} \left( \exp \left\{ - \frac{j^2}{2}
\int^{\frac{\epsilon^2 u}{4}}_{\frac{\epsilon^2 t}{4}} \xi(l) dl
\right\} \bigg| \xi \left( \frac{\epsilon^2 t}{4}  \right) , \xi
\left( \frac{\epsilon^2 u}{4} \right) \right)
\end{equation}
respectively. Regarding Eq. \eqref{eqcondLPtransform2}, we have the following result, see e.g. \cite{JeanblancYoCh09}.
\begin{eqnarray*}
\lefteqn{\tilde{E} \left( \exp \left\{ - \frac{b^2}{2} \int^t_0
\xi(s) ds \right\} \bigg| \xi(0) =x, \xi(t) =y  \right)}
\\ &=& \frac{bt}{sinh (bt)} \exp \left\{ \frac{x+y}{2t} (1 - b t coth (bt)) \right\} \frac{I_{\nu} ( b \sqrt{xy} /sinh (bt) ) }{I_{\nu}(\sqrt{xy}/t)} \, ,
\end{eqnarray*}
where $\nu=\frac{n}{2}-1$, the index of the squared Bessel process. Regarding Eq. \eqref{eqcondLPtransform2}, we proceed as follows:
\begin{eqnarray*}
\lefteqn{ \tilde{E} \left( \exp \left\{ - a \xi(t) - \frac{b^2}{2} \int^t_0 \xi(s) ds - c \int^t_0 \frac{ds}{\xi(s)} \right\} \right) }
\\ &=& \int^{\infty}_0 \exp \left\{ - a y \right\} \tilde{E} \left(  \exp \left\{ - \frac{b^2}{2} \int^t_0 \xi(s) ds - c \int^t_0 \frac{ds}{\xi(s)} \right\} \bigg| \xi(0)=x, \xi(t) = y \right)
\\ && q^{(\nu)}_t(x,y) dy \, ,
\end{eqnarray*}
where $q^{(\nu)}_t(x,y)$ denotes the transition density of $\xi$ started from $x$ at $0$ being in $y$ at time $t$. Consequently, we can look at
\begin{displaymath}
\tilde{E} \left( \exp \left\{ - a \xi(t) - \frac{b^2}{2} \int^t_0 \xi(s) ds - c \int^t_0 \frac{ds}{\xi(s)} \right\} \right)
\end{displaymath}
as the Laplace transform of
\begin{displaymath}
\tilde{E} \left(  \exp \left\{ - \frac{b^2}{2} \int^t_0 \xi(s) ds
- c \int^t_0 \frac{ds}{\xi(s)} \right\} \bigg| \xi(0)=x,  \xi(t) =
y \right) q^{(\nu)}_t(x,y) \, .
\end{displaymath}
This Laplace transform is explicitly inverted in \cite{CraddockLe09}, see Example 5.3:
\begin{eqnarray*}
\lefteqn{ \tilde{E} \left(  \exp \left\{ - \frac{b^2}{2} \int^t_0
\xi(s) ds - c \int^t_0 \frac{ds}{\xi(s)} \right\} \bigg| \xi(0) =
x, \xi(t) = y \right) q^{(\nu)}_t(x,y)}
\\ &=& \frac{b}{2 sinh (bt)} \exp \left\{ - b(x+y)/(2 tanh (bt)) \right\} \left( \frac{y}{x} \right)^{(n-2)/4} I_{\sqrt{(n-2)^2 + 8 c}/2} \left( \frac{b \sqrt{xy}}{sinh(b t)} \right) \, ,
\end{eqnarray*}
and
\begin{displaymath}
q^{(\nu)}_t(x,y) = \frac{1}{2 t} \left( \frac{y}{x} \right)^{\nu/2} \exp \left( - \frac{x+y}{2 t} \right) I_{\nu}(\frac{\sqrt{x y}}{t} ) \, ,
\end{displaymath}
where $\nu =n/2-1$. Hence we have
\begin{equation} \label{eqmomentgenfun1}
E \left( \exp \left\{ - a^* \int^u_t \frac{ds}{X_s}  \right\}
\bigg| X_t, X_u \right) = \frac{I_{\sqrt{\nu^2+8a/(\epsilon^2)}}
(\frac{j\sqrt{X_t X_u}}{\sinh (j \Delta)}) }{I_{\nu}(\frac{j
\sqrt{X_t X_u} }{\sinh ( j \Delta)})} \, ,
\end{equation}
where $\Delta=\frac{ u \epsilon^2}{4} - \frac{t \epsilon^2}{4}$.
We have characterized the conditional distribution of $\int^u_t
\frac{ds}{X_s} \bigg| X_t, X_u$. We will sample from this
distribution by inversion, which we discuss in Section
\ref{secimplementation}.

\section{Implementation} \label{secimplementation}

In this section, we discuss the implementation of Algorithm \ref{alg:3over2}, where we rely on \cite{BroadieKa06} and \cite{GlassermanKi11}. The implementation of Step 1) was already discussed in Section \ref{secMonteCarloalg}, and Steps 3) and 4) are trivial, so we focus on Step 2). We obtain the characteristic function of
\begin{displaymath}
\int^u_t \frac{ds}{X_s} \bigg| X_t \, , X_u
\end{displaymath}
by setting $a^*=- i a$ in \eqref{eqmomentgenfun1} and we use the notation
\begin{displaymath}
\Phi(a) = E \left( \exp \left\{ i a \int^u_t \frac{ds}{X_s}
\right\} \bigg| X_t, X_u \right)
\end{displaymath}
Furthermore, we define the conditional distribution function
\begin{eqnarray*}
F(x)&:=& P \left( \int^u_{t} \frac{ds}{X_s} \leq x \bigg| X_t, X_u
\right) =\frac{1}{\pi} \int^{\infty}_{=- \infty} \frac{\sin ( u
x)}{u} \Phi(u) du
\\ &=& \label{eqdefcondvar} \frac{2}{\pi} \int^{\infty}_0 \frac{sin(u x)}{u} \Phi(u) du \, ,
\end{eqnarray*}
where we do not emphasize the dependence of the cumulative distribution function on $X_t$ and $X_u$. We employ the trapezoidal rule to approximate the integral in Eq. \eqref{eqdefcondvar} numerically, and obtain
\begin{displaymath}
F(x) = \frac{h x}{\pi} + \frac{2}{\pi} \sum^{N}_{j=1} \frac{\sin (h j x)}{j} Re \left( \Phi( h j ) \right) - e_d(h) - e_T(N) \, ,
\end{displaymath}
where $h$ is the grid size associated with the trapezoidal rule and $N$ denotes the number of terms in the summation; we use $e_d(h)$ to denote the discretisation error associated with grid size $h$ and $e_T(N)$ denotes the truncation error resulting from the termination of the sum after $N$ terms. From the discussion in \cite{BroadieKa06}, it is known that choosing
\begin{displaymath}
h = \frac{2 \pi}{x + u_{\epsilon}} \, ,
\end{displaymath}
where $1- F(u_{\epsilon}) =\epsilon$ and $0 \leq x \leq u_{\epsilon}$, results in a discretization error $e_d(h)$ of magnitude $\epsilon$. However, as $u_{\epsilon}$ is difficult to obtain in this way, we set $u_{\epsilon}$ equal to the mean plus $12$ standard deviations. Regarding the value of $N$, as in \cite{BroadieKa06}, we terminate the sum at $j=N$, where
\begin{displaymath}
\frac{\vert \Phi(h N) \vert}{N} < \frac{\pi \epsilon}{2}
\end{displaymath}
and $\epsilon$ is the desired discretization error.

Having discussed the implementation of the conditional cumulative probability function, $F(x)$, we proceed as follows. Following \cite{GlassermanKi11}, Section 4.3, we evaluate $F( \cdot)$ on the grid
\begin{equation} \label{eqpointgrid}
x_i = w \mu + \frac{i-1}{M}(u_{\epsilon} - w \mu ) \, , i = 1, \dots, M+1 \, ,
\end{equation}
where $\mu$ denotes the conditional expected value of the integrated variance and
%\begin{displaymath}
%\mu = E \left( \int^u_t \frac{ds}{X_s} \bigg| X_t = x_1 , X_u = x_2 \right) \, ,
%\end{displaymath}
we choose $w=0.01$ and $M=200$. Consequently, we sample from the conditional distribution by inversion.

Clearly, the most time consuming step when sampling from the conditional distribution is the evaluation of the modified Bessel function of the first kind, $I_{v}(z)$, which has to evaluated at complex $v$. We point out that this operation is easily done in MATHEMATICA, and consequently we use the MATHEMATICA computing package to perform the simulation. We remark that we also perform the inversion of the probability distribution using MATHEMATICA, unlike \cite{BroadieKa06}, where the equation,
\begin{displaymath}
F(x) = U ,
\end{displaymath}
where $U$ is simulated from a uniform $[0,1]$ distribution, was solved for $x$ using Newton's method. The reason we do not employ their technique is twofold: Firstly, \cite{BroadieKa06} could use a good initial guess for Newton's method, as they could approximate the conditional distribution of the integrated variance by the Gaussian distribution. The analogous result for the $3/2$ model is not known, but we found that the Newton search is highly dependent on the initial guess. Secondly, using MATHEMATICA's built-in functions, we can perform this search both quickly and reliably, in particular, we compute the probability distribution on the grid of points \eqref{eqpointgrid}, and use MATHEMATICA's NEAREST function to identify the point at which the probability distribution assumes a value closest to the simulated uniform random variable. We point out that the computational time taken is highly dependent on one's experience with MATHEMATICA's procedural programming, and should the modified Bessel function of the first kind, allowing for $v$ to be complex, become available for other computing packages, or should one choose to implement it, the computational times can be expected to differ substantially. Hence we report the number of sample paths required to achieve a particular standard error. This metric is platform and user independent, so we find it useful to present. Furthermore, it ties in nicely with the variance reduction techniques presented in Section \ref{secvarred}, which allow us to identify the number of trajectories by which the computational effort is reduced. Finally, the Monte Carlo algorithm we present is of course parallelizable, in principle, one trajectory could be computed on one processor, which allows us to substantially reduce the computational effort, and we expect this trend to continue in the future.

We conclude this section by applying Algorithm \ref{alg:3over2} to the pricing of a European call option. This product is chosen, as it allows us to verify Algorithm \ref{alg:3over2} using a different method, namely we price the European call option via the characteristic function of the logarithm of $S_T$. The characteristic function of the logarithm of $S_T$ is well-known, see e.g. \cite{CarrSu07}, Theorem 3, or \cite{Heston97}, \cite{Lewis00}. We choose the following set of parameters
\begin{equation} \label{eqparscall}
S_0=1 \,  , \, K=1,  \, \kappa = 2 \, , \, \theta =1.5 \, , \,
\epsilon =0.2 \, , \, \rho = -0.5 \, , T=1 \, , r =0.05
\end{equation}
and obtain the reference value $0.443059$. Table \ref{tableMCcall} shows price estimates and standard errors for a given number of simulation trials.

\begin{table}
\begin{center}
\begin{tabular}{|c|c|c|}
\hline
\mbox{ Number of simulation trials } & \mbox{Price estimates } & \mbox{Standard error estimates} \\
\hline
2560 & 0.46787175 & 0.02721607 \\
\hline
10240 & 0.430909 & 0.0132623  \\
\hline
40960 & 0.442416 & 0.00672314  \\
\hline
\end{tabular}
\end{center}
\caption{Price estimates and standard error estimates for a European call option using Monte Carlo simulation}
\label{tableMCcall}
\end{table}

%\section{Numerical Examples}

\section{Variance Reduction Techniques for Stochastic Volatility Models} \label{secvarred}

We recall some well-known variance reduction techniques for stochastic volatility models, see in particular \cite{Willard97}. These methods are not restricted to the $3/2$ model, but are more generally applicable, see also \cite{BroadieKa06}. The key observation is that given paths of
\begin{displaymath}
\int^T_0 V_s ds \textrm{ and } \int^T_0 \sqrt{V_s} dW^1_s \, ,
\end{displaymath}
the price of a European call option is given by the Black-Scholes price, with modified initial share price
\begin{displaymath}
\tilde{S}_0 = S_0 \exp \left\{ - \frac{\rho^2}{2} \int^T_0 V_S ds + \rho \int^T_0 \sqrt{V_s} dW^1_s \right\} \, ,
\end{displaymath}
and adjusted volatility $\tilde{\sigma} \sqrt{1 - \rho^2}$, where
\begin{displaymath}
\tilde{\sigma} = \sqrt{\frac{1}{T} \int^T_0 V_s ds} \, .
\end{displaymath}
Consequently, using $BS \left( S_0, K, r , \tau, \sigma \right)$ to denote the Black-Scholes price of a European call with initial stock price $S_0$, strike $K$, interest rate $r$, time to maturity $\tau$, and volatility $\sigma$, we have
\begin{equation}
E \left( \exp \left\{ - r T \right\} \left( S_T - K \right)^+ \right) = E \left(  BS \left( \tilde{S}_0, K , r , \tau , \tilde{\sigma} \right) \right)  \, ,
\end{equation}
that is, we firstly simulate $\int^T_0 V_S ds$ and $\int^T_0 \sqrt{V_s} dW^1_s$, using Algorithm \ref{alg:3over2}, and then compute the Black-Scholes price, for the particular values of $\tilde{S}_0$ and $\tilde{\sigma}$ corresponding to the trajectory of $\int^T_0 V_s ds$ and $\int^T_0 \sqrt{V_s} dW^1_s$. This can of course be expected to reduce the variance, essentially, we do not estimate the Black-Scholes price using Monte Carlo simulation, which is done in Step 4 of Algorithm \ref{alg:3over2}, but compute the value exactly. Finally, it can be expected that when combining the conditional Monte Carlo approach with quasi-Monte Carlo points, the approach is even more efficient, see \cite{Willard97}. This is due to the fact that taking the conditional expectation has a smoothing effect, which can be expected to improve the performance of quasi-Monte Carlo methods, see \cite{LEcuyerLe00}, Subsection 10.1.

Regarding the quasi-Monte Carlo point sets, we employ the two-dimensional Sobol sequence, which is well-known to have the optimal quality parameter $t=0$. To be more precise, we use two-dimensional Sobol nets, comprised of $2^m$ points, where $m=5,6,7,$ and $8$, using the first coordinate to generate $X_T$ and the second coordinate to generate the conditional integrated variance. Furthermore, we randomize the nets using Owen's scrambling algorithm, \cite{Owen95}, implemented using the algorithm presented in \cite{HongHi03}: we produce $30$ independent copies of each net, allowing us to estimate standard errors. Finally, we remark that besides the ability to estimate standard errors, the randomized point sets can also be expected to produce better convergence rates, see \cite{DickPi10}, \cite{Owen97}.

We use the set of parameters \eqref{eqparscall} and, in Table \ref{tableqMCcall}, we report estimates of the option price and standard errors. The number of simulation trials performed is $30*2^m$, where $m=5,6,7,$ and $8$, as we subject each Sobol net to $30$ randomizations.

\begin{table}
\begin{center}
\begin{tabular}{|c|c|c|}
\hline
\mbox{ Number of simulation trials } & \mbox{Price estimates} & \mbox{Standard error estimates} \\
\hline
960 & 0.441047 & 0.00286267 \\
\hline
1920 & 0.443279 & 0.00195695  \\
\hline
3840 & 0.442995 & 0.000698016  \\
\hline
7680 & 0.44392 & 0.000461543  \\
\hline
\end{tabular}
\end{center}
\caption{Price estimates and standard error estimates for a European call option using quasi-Monte Carlo points}
\label{tableqMCcall}
\end{table}

We note that conditional Monte Carlo combined with quasi-Monte Carlo point sets provides us with a substantial variance reduction. Exploring similar variance reduction techniques applicable to other payoff functions poses interesting and important future research questions.

\section{Conclusion} \label{secconc}

In the present paper, we provided an exact simulation algorithm for the $3/2$ model. A result by Craddock and Lennox allowed us to adapt the Broadie-Kaya algorithm for affine processes to the $3/2$ model. Furthermore, we discussed variance reduction techniques and found that conditional Monte Carlo combined with quasi-Monte Carlo point sets provided significant variance reduction.

In future work, we aim to discuss path-dependent payoffs under the $3/2$ model and provide effective variance reduction techniques for path-dependent payoffs.

\setlinespacing{1.}

%\bibliographystyle{\dir chicago}

%\bibliographystyle{dcu_ic_k}
%\bibliography{\dir my}
%\bibliography{\dir my}

\begin{thebibliography}{10}
\bibitem{AhnGa99} Ahn, D.-H., and Gao, B., A Parametric Nonlinear Model of Term Structure Dynamics, Rev. Financial Studies, 12, 721--762, 1999.
\bibitem{BeskosRo05} Beskos, A., Papaspiliopoulos, O., and Roberts, G., Retrospective exact simulation of diffusion sample paths with applications, Bernoulli, 12, 1077--1098, 2006.
\bibitem{BeskosPaRo06} Beskos, A., Papaspiliopoulos, O., and Roberts, G., A factorisation of diffusion measure and finite sample path constructions, Methodology and Comp. Appl. Prob., 10, 85--104, 2008.
\bibitem{BeskosPaRo08} Beskos, A., and Roberts, G., Exact simulation of diffusions, Ann. Appl. Prob., 15, 2422--2444, 2008.
\bibitem{BroadieKa06} Broadie, M., and Kaya, O., Exact simulation of stochastic volatility and other affine jump diffusion processes, Oper. Res., 54, 217--231, 2006.
\bibitem{CarrSu07} Carr, P., and Sun, J., A new approach for option pricing under stochastic volatility, Rev. Derivatives. Res., 10, 87--150, 2007.
\bibitem{Chen08} Chen, N., Exact simulation of stochastic differential equations, Chinese Univ. of Hong Kong (working paper).
\bibitem{CraddockLe07} Craddock, M., and Lennox, K., Lie group symmetries as integral transforms of fundamental solutions, J. Differential Equations, 232, 652--674, 2007.
\bibitem{CraddockLe09} Craddock, M., and Lennox, K., The calculation of expectations for classes of diffusion processes by Lie symmetry methods, Ann. Appl. Probab., 19, 127--157, 2009.
\bibitem{CreutzigDeMuRi09} Creutzig. J., Dereich, S., M\"uller-Gronbach, T., and Ritter, K., Infinite-dimensional quadrature and approximation of distributions, Foundations of Computational Mathematics, 9, 391--429, 2009.
\bibitem{DickPi10} Dick, J., and Pillichshammer, F., Digital Nets and Sequences. Discrepancy Theory and Quasi-Monte carlo Integration, Cambridge Unievrsity Press, 2010.
\bibitem{DuffieGl95} Duffie, D., and Glynn, P. W., Efficient Monte Carlo simulation of security prices, Ann. Appl. Probab., 5, 897--905, 1995.
\bibitem{Giles08} Giles, M. B., Multi-level Monte Carlo path simulation, Oper. Res., 56, 607--617, 2008.
\bibitem{GlassermanKi11} Glasserman, P., and Kim, K.-K., Gamma expansion of the Heston stochastic volatility model, Finance Stoch., 15, 267--296, 2011.
\bibitem{Heston93} Heston, S. L., A closed-form solution for options with stochastic volatility with applications to bond and currency options, Rev. Financial Studies, 6, 327--343, 1993.
\bibitem{Heston97} Heston, S. L., A simple new formula for options with stochastic volatility, Washington University of St. Louis (working paper).
\bibitem{HongHi03} Hong, H., and Hickernell, F., Algorithm 823: Implementing scrambled digital sequences, ACM Transactions on Mathematical Software, 29, 95--109, 2003.
\bibitem{ItkinCa10} Itkin, A., and Carr, P., Pricing swaps and options on quadratic variation under stochastic time change models - discrete observations case, Rev. Derivatives Res., 13, 141--176, 2010.
\bibitem{JeanblancYoCh09} Jeanblanc, M., Yor, M., and Chesney, M., Mathematical Methods for Financial Markets, Springer Finance, Springer, 2009.
\bibitem{KloedenPl95} Kloeden, P. E., and Platen, E., Numerical Solution of Stochastic Differential Equations, Springer, 1999.
\bibitem{LaurenceWa05} Laurence, P., and Wang, T.-H., Closed form solutions for quadratic and inverse quadratic term structure models, Int. J. Theor. Appl. Finance, 8, 1059--1083, 2005.
\bibitem{LEcuyerLe00} L'Ecuyer, P., and Lemieux, C., Variance Reduction via Lattice Rules, Management Science, 46, 1214--1235, 2000.
\bibitem{Lewis00} Lewis, A., L., Option Valuation Under Stochastic Volatility, Finance Press, Newport Beach, 2000.
\bibitem{Owen95}Owen, A. B., Randomly permuted $(t,m,s)$-nets and $(t,s)$-sequences. In H. Niederreiter and J.-S. Spanier (Eds.), Monte Carlo and quasi-Monte Carlo methods in scientific computing, 299--317, Springer, 1995.
\bibitem{Owen97} Owen, A. B., Monte Carlo variance of scrambled quadrature, SIAM J. Numer. Anal., 34, 1884--1910, 1997.
\bibitem{PitmanYo82} Pitman, J., and Yor, M., A decomposition of Bessel bridges, Probab. Theory Related Fields, 59, 425--457, 1982.
\bibitem{PlatenHe06} Platen, E., and Heath, D., A Benchmark Approach to Quantitative Finance, Springer Finance, Springer, 2006.
\bibitem{RevuzYo99} Revuz, D., and Yor, M., Continuous Martingales and Brownian Motion, 3rd edition, Springer, 1999.
\bibitem{Willard97} Willard, G. A., Calculating prices and sensitivities for path-independent derivative securities in multi-factor models, J. Derivatives, 5, 45--61, 1997.
\end{thebibliography}

\end{document}